\documentclass{cjaa}                   % preprint, the final version for publication
\usepackage{graphicx}                   %for PS/EPS graphics inclusion, new
\input{epsf.sty}                        %for PS/EPS graphics inclusion, old
\input{psfig.sty}                       %for PS/EPS graphics inclusion, old
\begin{document}

   \title{Astrophysical Quark Matter}
   \volnopage{Vol.0 (200x) No.0, 000--000}      %%preserved for Editor. DOn't remove!
   \setcounter{page}{1}          %%starting page, preserved for Editor. DOn't remove!

   \author{Renxin Xu}
   \offprints{Renxin Xu}
   \institute{School of Physics, Peking University,
             Beijing 100871, China\\
             \email{rxxu@bac.pku.edu.cn}
             }

   \date{Received~~~~~~~~~~~; ~~~accepted}

   \abstract{
The quark matter may have great implications in astrophysical
studies, which could appear in the early Universe, in compact
stars, and/or as cosmic rays. After a general review of
astrophysical quark matter, the density-dominated quark matter is
focused.
   }

   \authorrunning{Renxin Xu}            %author_head in even pages
   \titlerunning{Astrophysical quark matter}  % title_head in odd pages

   \maketitle

\section{Introduction to quark matter}

{\em Twenty years ago, a paper on astrophysical quark matter was
published (Witten~\cite{Witten84}), of which I will be in memory
here}. Anyway, let's begin with the elements of particle physics.

One of the most great achievements in the 20th century is the
construction of the standard model in particle physics, which
asserts that the material in the universe is made up of elementary
fermions (divided into leptons and quarks) interacting though the
Yang-Mills gauge bosons.
The fermions have three generations (1st: \{$\nu_{\rm e}$, e; u,
d\}, 2nd: \{$\nu_{\rm \mu}$, $\mu$; c, s\}, and 3rd: \{$\nu_{\rm
\tau}$, $\tau$; t, b\}). The six flavors of quarks can be
classified into light (up $u$, down $d$, and strange $s$) and
heavy (charm $c$, top $t$, and bottom $b$) quarks according to
their masses, and each quark is charged by one of the three
colors.
The elemental interactions satisfy the local gauge symmetries
(sometimes to be broken spontaneously), which result in different
interaction bosons: photon (electromagnetic), W$^\pm$ and Z$^0$
(weak), 8 types of gluons (strong), and probably graviton
(gravitational).
The gauge theory for electromagnetic and weak interactions is very
successful, with a much high precision in calculation.
Additionally, one can also apply effectively Einstein's general
theory of relativity to deal with gravitational phenomena if the
scale is not as small as the Plank scale ($\sim 10^{-33}$ cm or
$\sim 10^{28}$ eV) although a gauge theory of gravity is still not
available.
However, as for the gauge theory for strong interaction, the
quantum chromodynamics (QCD) is still developing, into which many
particle physicists are trying to make efforts.

Nonetheless, QCD has two general properties. For strong
interaction in small scale ($\sim 0.1$ fm), i.e., in the high
energy limit, the interacting particles can be treated as being
{\em asymptotically free}; a perturbation theory of QCD (pQCD) is
possible in this case.
Whereas in larger scale ($\sim 1$ fm), i.e., in the low energy
limit, the interaction is very strong, which results in {\em color
confinement}. The pQCD is not applicable in this scale (many
non-perturbative effects appear then), and a strong interaction
system can be treated as a system of hadrons in which quarks and
gluons are confined. In this limit, we still have effective means
to study color interaction: 1, the lattice formulation (LQCD),
with the discretization of space-time and on the base of QCD,
provides a non-perturbative framework to compute numerically
relations between parameters of the standard model and
experimental phenomena; 2, phenomenological models, which rely on
experimental date available at low energy density, are advanced
for superdense hadronic and/or quark matter.

These two features result in two distinct {\em phases}
(Rho~\cite{Rho01}) of color-interacting system, depicted in the
QCD phase diagram in terms of temperature $T$ vs. baryon chemical
potential $\mu_{\rm B}$ (or baryon number density).
Hadron matter phase locates at the low energy-density limit where
both $T$ {\em and} $\mu_{\rm B}$ are relatively low, while a new
phase called {\em quark gluon plasma} (QGP) or {\em quark matter}
appears in the other limit when $T$ ({\em or} $\mu_{\rm B}$) is
high although this new state of matter is still not found with
certainty yet.
It is therefore expected that there is a kind of phase transition
from hadron gas to QGP (or reverse) at critical values of $T$ and
$\mu_{\rm B}$.
Actually a deconfinement transition is observed in numerical
simulations of LQCD for zero chemical potential $\mu_{\rm B}=0$,
when $T\rightarrow T_{\rm c}\simeq (150\sim 180)$ MeV.

Physics is essentially an experimental science. Can we find quark
matter in reality? Certainly we may improve substantially the
knowledge about the strong interaction by studying the matter's
various properties if a QGP state is identified in hand.
One way is to create high energy-density fireball in laboratory
through the collisions of relativistic heavy ions in accelerators.
Quark matter is expected in the center of the fireball with a
temperature of $T_{\rm c}$, but the QGP is hadronized soon and it
is the hadronic matter to be detected in the final states.
It is then a challenge to find clear signatures of QGP without
ambiguousness in the terrestrial laboratory physics (lab-physics).
However, it may be a {\em shortcut} to study astrophysical quark
matter, since astrophysics offers an alternative channel for us to
explore the fundamental laws in the nature.

In fact, there could be three scenarios for the existence of
astrophysical quark matter (Witten \cite{Witten84}).
1. The temperature and density of the universe decrease in the
standard model of cosmology, and it is therefore expected various
of phase-transitions would occur in the early universe. One of
such transitions could be \{QGP $\rightarrow$ hadron gas\} (called
quark-hadron phase transition) when cosmic temperature was $T\sim
200$ MeV.
2. Quark stars is another possibility. It is conventionally
believed that pulsar-like stars, which are the residua of
supernova explorations of evolved massive stars, are normal
neutron stars (Lattimer \& Prakash~\cite{lp04}) composed
dominantly of neutron fluid. However, recent observations indicate
that such pulsar-like stars could actually be quark stars composed
of quark matter. One kind of the quark stars are those with
strangeness, which can be considered to be most stable bulk quark
matter (with nearly equal number of $u$, $d$, and $s$ quarks).
Such quark stars are named as strange quark stars, or simply
strange stars (Xu~\cite{Xu04}), because they have $s$ (strange)
quarks discovered in 1940s.
3. It is an alternative possibility that quark matter with baryon
numbers $>10^3$ could appear in cosmic rays. Recent discovery of
ultra-high energy cosmic rays may be a hint, since the energy of
quark nuggets could be greater than the GZK cutoff (Madsen \&
Larsen ~\cite{mad03}, Xu \& Wu~\cite{xw03}).

There are two {\em different} kinds of quark matter to be
investigated in lab-physics and astrophysics, which appear in two
regions in the QCD phase diagram.
Quark matter in lab-physics and in the early universe is
temperature-dominated ($T\ll 0, \mu_{\rm B}\sim 0$), while that in
quark stars or as cosmic rays is density-dominated ($T\sim 0,
\mu_{\rm B}\ll 0$). (Experiments for \{$T\sim 0, \mu_{\rm B}\gg
0$\} may be possible in the future.)
Previously, Monte Carlo simulations of LQCD were only applicable
for cases with $\mu_{\rm B}=0$. Only recent attempts are tried at
$\mu_{\rm B}\neq 0$ (quark stars or nuggets) in LQCD. We have then
to rely on phenomenological models to speculate on the properties
of density-dominated quark matter by extrapolating our knowledge
at nuclear matter density.

\section{Quark-hadron transition in the early universe}

The underlying gauge symmetry in high-temperature is much larger
than that in low-$T$, the vacuum may then undergo various
phase-transitions in case of symmetry broking spontaneously when
the universe cooled as it expanded. Such transitions would result
in the remnants of fase vacuum: topologically defects
(0D-monopoles, 1D-cosmic strings, and 2D-domain walls), and may
even induce the universe to inflate (Kolb \& Turner~\cite{kt90}).
In the early radiation-dominated universe, whose space-time is
described by Robertson-Walker metric, the temperature can be
approximated simply by
\begin{equation}
T\sim {1~{\rm MeV}\over \sqrt{t}},
\label{Tff}
\end{equation}
with the cosmic age $t$ in unit of seconds.
A electro-weak transition occurred at temperature $T_{\rm ew}\sim
100-200$ GeV when the unverse aged $t_{\rm ew}\sim 10^{-11}$ s,
while a quark-hadron phase-transition (QHPT, or QCD transition)
took place at temperature $T_{\rm qcd}\sim 100-200$ MeV when the
cosmic age was $t_{\rm qcd}\sim 10^{-5}$ s (see a review of, e.g.,
Schwarz~\cite{sch03}).

The cosmic QHPT is very close to an equilibrium process, since the
the relaxation time scale of color interaction, $\sim 1~{\rm
fm}/c\sim 10^{-23}$ s, is much smaller than the time interval
$t_{\rm qcd}\sim 10^{-5}$ s in which the cosmic thermodynamical
variables and expanding-dynamical curvature can change
significantly.
A first-order (or second-order, actually the order is still not
certain) QCD transition may proceed through bubble nucleation. The
hadronic bubbles grow, release laten energy, and could collide
with others when they are larger enough (i.e., bubbles with hadron
gas grew until they merge and filled up the whole universe in the
end of QHPT).
The horizon radius at that time is $R_{\rm h}\sim ct_{\rm qcd}\sim
10$ km. However, the typical separation between bubbles, $D_{\rm
b}$, could be much smaller than the horizon radius, only $D_{\rm
b}\sim 10^{-6}R_{\rm h}\sim 1$ cm according to lattice QCD
calculations where the bubble surface tension and laten heat are
included.

The cosmic QHPT may have many astrophysical consequences which
would test the physical process in turn.
Big-bang nucleosynthesis (BBN) predicts the abundances of the
light elements (D, $^3$He, $^4$He, and $^7$Li) synthesized at
cosmic age of $\sim 10^3$ s, which are observation-tested spanning
{\em nine} orders of magnitude (number ratios: from $^4$He/H$\sim
0.08$ down to $^7$Li/H $\sim 10^{-10}$).
However, the initial physical conditions for BBN should be setted
by cosmic QHPT. For instance, the inhomogeneities of temperature
and baryon numbers during bubble nucleation may affect the
abundances synthesized, which may clear the possible inconsistency
of the light element abundances with the CMB measurements (Cyburt,
Fields \& Olive~\cite{cfo03}).
In this sense, BBN offers then a reliable probe of QHPT. As a
result, this study could provide a better determination of the
baryonic density in the universe.

The formation of quark nuggets could be another probable
consequence.
Towards the end of the QHPT, baryon-enriched quark droplets
shrank, and might remain finally to play the role of dark matter
(Witten~\cite{Witten84}). Quark droplets with strangeness are
conjectured to absolutely stable (Bodmer~\cite{bod71},
Witten~\cite{Witten84}), and the residual quark nuggets could then
probably be composed of strange quark matter with high baryon
density.
Can we detect such quark nuggets? Two candidates in reality: the
ultra-high energy cosmic rays (UHECRs) with energy beyond the GZK
cutoff and the massive compact halo objects (MACHOs) discovered
through gravitational microlensing (Alcock et
al.~\cite{alcock93}).
It is worth noting, that MACHOs could be probably low-mass quark
stars formed by evolved stars, rather than quark nuggets born
during the QHPT (Banerjee et al.~\cite{ban03}) if pulsar-like
stars is actually quark stars. Beside, strangelets can also form
through stellar process.
Additionally, the relic quark-nuggets may evaporate baryons as
they cool, and they can hardly exist today (Bhattacharjee et
al.~\cite{bha93}).
In conclusion, the QHPT quark nuggets could not be very necessary
to understand the observations, although the existence possibility
of which can not be ruled out.

There could be other relics of cosmic QHPT. An very interesting
issue is to study the bubble collisions which may be responsible
to the generation of gravitational waves (Witten~\cite{Witten84})
as well as the large-scale magnetic walls which may lead to
observable polarization correlations and density fluctuations in
cosmic microwave background radiation (Kisslinger~\cite{ki02}).
Seed magnetic fields could be produced by currents on the bubble
surface (e.g., Hindmarsh \& Everett~\cite{he98}).

\section{Quark matter in compact stars and as cosmic rays}

In different locations of the diagram (Fig. 1), besides that the
interaction strength between quarks and gluons is weak or strong,
the vacuum would have different features and is thus classified
into two types: the perturbative-QCD (pQCD) vacuum and
nonperturbative-QCD (QCD) vacuum. The coupling is weak in the
former, but is strong in the later.
Quark-antiquark (and gluons) condensations occur in QCD vacuum
(i.e., the expected value of $\langle {\bar q}q\rangle \neq 0$),
but not in pQCD vacuum.
The chiral symmetry is spontaneously broken in case the vacuum is
changed from pQCD to QCD vacuums, and quarks become then massive
constituent ones.
LQCD calculations (Kogut~\cite{kog91}) show that the value of
$\langle {\bar q}q\rangle$ increases when the color coupling
becomes strong (i.e., temperature or baryon density decrease).
Therefore, we note that the quark de-confinement and the chiral
symmetry restoration may {\em not} take place at a same time.
%------------------- Fig1: QCD phase diagram ---------------------
\begin{figure}[t]
  \centering
  \begin{minipage}[t]{.5\textwidth}
    \centering
    \includegraphics[height=8cm,width=10cm]{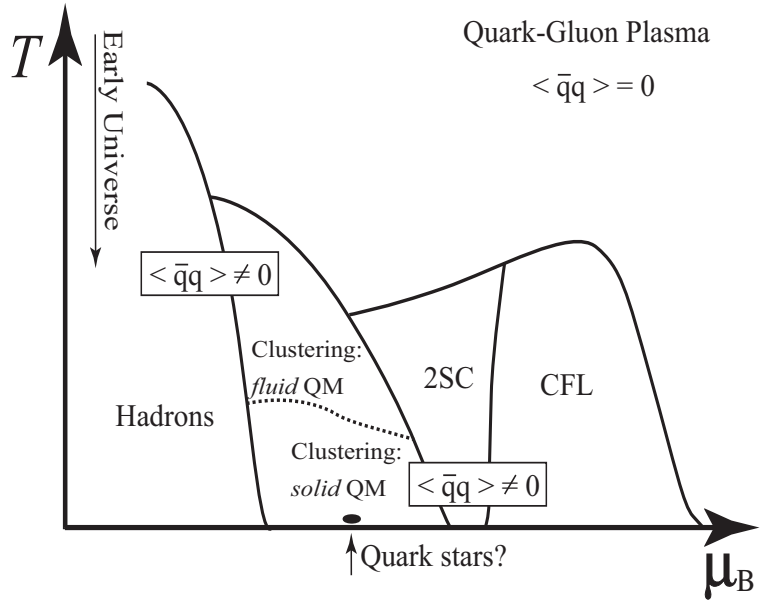}
  \end{minipage}
     \caption{
Schematic illustration of QCD phase diagram.
              }
\end{figure}

Considerable theoretical efforts have been made in past years to
explore the QCD phase diagram.
When $T$ or $\mu_{\rm B}$ are extremely high, there should be QGP
phase because of the asymptotic freedom, and vacuum is of pQCD.
However, in a relatively lower energy limit, especially in the
density-dominated region, the vacuum is phase-converted to QCD one
but the quarks could be still deconfined. It is a hot point to
investigate the possibility that real quarks may also be condensed
(i.e., $\langle q q\rangle \neq 0$) simultaneously when $\langle
{\bar q}q\rangle \neq 0$, the so-called color-superconducting
(CSC) phases (for recent reviews, see, e.g., Ren~\cite{ren04},
Rischke~\cite{ris04}). Actually two CSC phases are currently
discussed. One corresponds to Cooper pairing among the two flavors
of quarks ($u$ and $d$) only, the two-flavor color
superconductivity (2SC) phase, in case that $s$ quark is too
massive to participate. Another one occurs at higher $\mu_{\rm B}$
in which $s$ quarks are relatively less massive and are thus
involved in Cooper pairing, the color-flavor locked (CFL) phase.

However, another possibility can not be ruled out:  $\langle q
q\rangle = 0$ while $\langle {\bar q}q\rangle \neq 0$.
When $T$ is not high, along the reverse direction of the $\mu_{\rm
B}$ axis, the value of $\langle {\bar q}q\rangle$ increases, and
color coupling between quarks and gluons becomes stronger and
stronger.
The much strong coupling may favor the formation of $n-$quark
clusters ($n$: the number of quarks in a cluster) in the case
(Xu~\cite{xu03}).
Recent experimental evidence for multi-quark ($n > 3$) hadrons may
increase the possibility of quark clustering.
The clusters are localized\footnote{
We apply ``local'' to refer that ``quark wavefunctions do almost
not overlap''. In this sense, localized clusters can still move
from place to place when $T$ is high, but could be solidified at
low $T$.
} %
to become {\em classical} (rather than quantum) particles when the
thermal de Broglie wavelength of clusters $\lambda\sim
h/\sqrt{3mkT}<l\sim [3n/(4\pi fn_{\rm b})]^{1/3}$ ($m$: the mass
of clusters, $l$: the mean cluster distance, $n_{\rm b}$: the
baryon number density, $f$: quark flavor number), assuming no
interaction is between the clusters.
Calculation based on this inequality is shown in Fig. 2 for the
case with strangeness ($f=3$). One sees that cluster localization
still exists even in very low temperature if $n\sim 10^2$. In
addition, the interaction in-between, which is neglected in the
calculation of Fig. 2, would also favor this localization.
%
%------------------- Fig2: The localization condition ---------------------
\begin{figure}[t]
  \centering
  \begin{minipage}[t]{.5\textwidth}
    \centering
    \includegraphics[height=8cm,width=10cm]{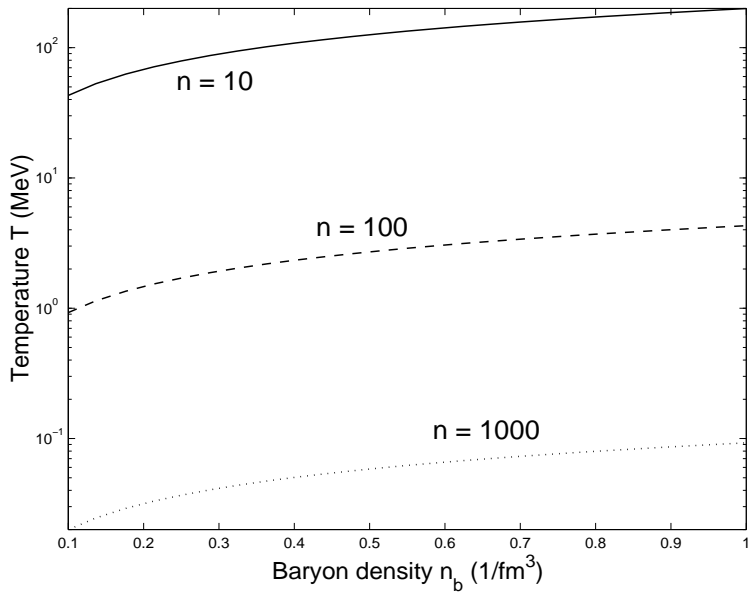}
  \end{minipage}
     \caption{
The conditions that lead to quark clusters being localized in
strange quark matter for different $n$. The condition may be
satisfied above a corresponding line. The mass of an $n-$quark
cluster is $300n$ MeV here. Note: the nuclear saturation density
is 0.16 fm$^{-3}$.
              }
\end{figure}

In case of negligible interaction, quark cluster would become
condensed when the temperature is lower than that shown in Fig. 2.
However, the interaction is certainly not weak since the vacuum is
of QCD ($\langle {\bar q} q \rangle\neq 0$).
Now, a {\em competition} between condensation and solidification
appears then, just like the case of laboratory low-temperature
physics.
Quark matter would be solidified as long as the interaction energy
between neighboring clusters is much larger than that of the
kinetic thermal energy.
This is why {\em only helium}, of all the elements, shows
superfluid phenomenon though other noble elements have similar
weak strength of interaction due to filled crusts of electrons.
The essential reason for the occurrence of CSC is that there is an
attractive interaction between two quarks at the Fermi surface.
But, as discussed, much strong interaction may result in the quark
clustering and in a solid state of quark matter.
In conclusion, a new phase with $\langle {\bar q} q\rangle\neq 0$
but $\langle qq \rangle=0$ is suggested to be inserted in the QCD
phase diagram (Fig. 2).

Nevertheless, what can we know from experiments in case of those
theoretical uncertainties?

Astrophysics may {\em teach} us about the nature of
density-dominated quark matter, since experiments (in low-energy
heavy ion colliders) with low temperature but high density is only
possible in the future. Two scenarios of such astrophysical quark
matter: quark stars and ultra-high energy cosmic rays.
Fortunately and excitedly, many astrophysical challenges could not
exist any more in the solid quark star model for pulsar-like
stars, e.g., the thermal X-ray spectra (Xu~\cite{xu03}) and the
discrepancy between glitches and free-precessions (Zhou et
al.~\cite{z04}).

\section{Conclusions}

Possible astrophysical quark matter in three scenarios is
explained. The nature of density-dominated quark matter could be
uncovered by the study of pulsar-like stars, while it is still not
sure to link astrophysical observations to cosmic QCD phase
transition.

\begin{acknowledgements}
This work was funded by the Natural Science Foundation of China
(NSFC No. 10273001) and the Special Funds for Major State Basic
Research Projects of China (G2000077602). I am indebted to Dr. P.
F. Zhuang for a discussion of QCD phase diagram.
\end{acknowledgements}

\label{lastpage}

\end{document}